\documentclass[12pt]{article}
\usepackage{amsmath}
\usepackage{amsfonts}   
\usepackage{amssymb}    
\usepackage{graphicx}   
\usepackage{xcolor}
\usepackage{bm}
\usepackage{secdot}		
\usepackage{mathptmx}
\usepackage{siunitx}
\usepackage[version=4]{mhchem}
\usepackage{float}
\usepackage[utf8]{inputenc}
\usepackage{textcomp}
\usepackage[hang,flushmargin,bottom]{footmisc} 
\usepackage{authblk}
\usepackage{pdfpages}
\usepackage{titlesec}
\usepackage{enumitem}
\titleformat{\section}{\normalsize\bfseries}{\thesection.}{1em}{}	
\titleformat*{\subsection}{\normalsize\bfseries}

\usepackage{tocloft}	

\usepackage{enumitem}         

\usepackage[numbers,sort&compress]{natbib} 

\usepackage[hidelinks]{hyperref}
\hypersetup{
	colorlinks = true,
urlcolor ={blue},
citecolor = {.},
linkcolor = {.},
anchorcolor = {.},
filecolor = {.},
menucolor = {.},
runcolor = {.}
pdftitle={},
pdfsubject={},
pdfauthor={}, 
pdfkeywords={} 
}
\usepackage{epstopdf} 

\usepackage{fancyhdr, lastpage}	
\usepackage{caption} 
\newcommand{\e}{e$^-$}

\newcommand{\trit}{$^3$H}
\newcommand{\ber}{$^7$Be}
\newcommand{\sod}{$^{22}$Na}
\newcommand{\pbten}{$^{210}$Pb}
\newcommand{\isotope}[2]{\ce{^{#2}{#1}}}

\titlespacing*{\paragraph}  {0pt}{0pt}{.5em}

\begin{document}
	\urlstyle{rm} 
	
\vfill

\title{The Oscura Experiment}

\author[1]{Alexis Aguilar-Arevalo}
\author[2]{Fabricio Alcalde Bessia}
\author[2]{Nicolas Avalos}
\author[4]{Daniel Baxter}
\author[2]{Xavier Bertou}
\author[3]{Carla Bonifazi}
\author[4]{Ana Botti}
\author[5]{Mariano Cababie}
\author[4]{Gustavo Cancelo}
\author[1]{Brenda Aurea Cervantes-Vergara}
\author[6]{Nuria Castello-Mor}
\author[7]{Alvaro Chavarria}
\author[4,8]{Claudio R. Chavez}
\author[8]{Fernando Chierchie}
\author[9]{Juan Manuel De Egea}
\author[1]{Juan Carlos D'Olivo}
\author[10,15]{Cyrus E.~Dreyer}
\author[4,14]{Alex Drlica-Wagner}
\author[10]{Rouven Essig}
\author[4]{Juan Estrada}
\author[2]{Ezequiel Estrada}
\author[11]{Erez Etzion}
\author[4]{Guillermo Fernandez-Moroni}
\author[10]{Marivi Fern\'andez-Serra}
\author[12]{Steve Holland}
\author[6]{Agustin Lantero Barreda}
\author[4]{Andrew Lathrop}
\author[2]{Jos\'{e} Lipovetzky}
\author[13]{Ben Loer}
\author[14,4]{Edgar Marrufo Villalpando}
\author[9]{Jorge Molina}
\author[14]{Sravan Munagavalasa}
\author[14]{Danielle Norcini}
\author[5]{Santiago Perez}
\author[14]{Paolo Privitera}
\author[5]{Dario Rodrigues}
\author[13]{Richard Saldanha}
\author[9]{Diego Santa Cruz}
\author[10]{Aman Singal}
\author[4]{Nathan Saffold}
\author[4]{Leandro Stefanazzi}
\author[2]{Miguel Sofo-Haro}
\author[4]{Javier Tiffenberg}
\author[9]{Christian Torres}
\author[4]{Sho Uemura}
\author[6]{Rocio Vilar}

\affil[1]{Universidad Nacional Aut\'onoma de M\'exico, Ciudad de M\'exico, M\'exico}
\affil[2]{Centro Atomico Bariloche, Rio Negro, Argentina}
\affil[3]{International Center of Advanced Studies and Instituto de Ciencias Físicas, ECyT-UNSAM
and CONICET, Argentina }
\affil[4]{Fermilab National Accelerator Laboratory, IL, USA}
\affil[5]{Universidad de Buenos Aires, Buenos Aires, Argentina}
\affil[6]{Instituto de F\'isica de Cantabria (IFCA), CSIC–Universidad de Cantabria, Santander, Spain }
\affil[7]{University of Washington, WA, USA}
\affil[8]{IIIE CONICET and DIEC Universidad Nacional del Sur, Argentina}
\affil[9]{Facultad de Ingenier\'ia, Universidad Nacional de Asunci\'on, Paraguay}
\affil[10]{Stony Brook University, NY, USA}
\affil[11]{Tel Aviv University, Israel}
\affil[12]{Lawrence Berkeley National Laboratory, CA, USA}
\affil[13]{Pacific Northwest National Laboratory, WA, USA}

\affil[14]{University of Chicago, IL, USA}
\affil[15]{Center for Computational Quantum Physics, Flatiron Institute,  NY, USA}

\maketitle
\thispagestyle{empty}

\section*{Abstract}
The Oscura experiment will lead the search for low-mass dark matter particles using a very large array of novel silicon Charge Coupled Devices  (CCDs) with a threshold of two electrons and with a total exposure of 30 kg-yr. The R\&D effort, which began in FY20,
is currently entering the design phase with the goal of being ready to start construction in late 2024. Oscura will have unprecedented sensitivity to sub-GeV dark matter particles that interact with electrons, probing dark matter-electron scattering for masses down to $\sim$500~keV and dark matter being absorbed by electrons for masses down to $\sim$1~eV. 
The Oscura R\&D effort has made some significant progress on the main technical challenges of the experiment, of which the most significant are engaging new foundries for the fabrication of the CCD sensors, developing a cold readout solution, and understanding the experimental backgrounds. 

\begin{center}
	\tableofcontents
\end{center}


\pagebreak

\clearpage
\pagenumbering{arabic}

\section{Scientific Background and Goals of the Oscura Experiment}
\label{sec:ccience}

Identifying the nature of dark matter (DM) is one of the most important missions of particle physics and astrophysics today, and direct-detection experiments play an essential role in this endeavor.  The search for DM particles with masses up to a few orders of magnitude below the proton mass (``sub-GeV DM'') represents an important new experimental frontier that has been receiving increased attention, e.g.~\cite{Essig:2011nj,Essig:2013lka,Alexander:2016aln,Battaglieri:2017aum,BRNreport}.

Traditional direct-detection searches, which look for DM particles scattering elastically off of nuclei, typically have very little sensitivity to sub-GeV mass DM.  Indeed, the best current bounds below 1~GeV are limited to very large cross sections and are absent below $\sim$87~MeV~\cite{Abdelhameed:2019hmk,Angloher:2017sxg,Petricca:2017zdp,Agnese:2015nto,Alkhatib:2020slm}. While improved sensitivity at low DM masses to nuclear recoils from {\it elastic} scatters is possible with the next generation of experiments (see e.g.~\cite{BRNreport,SPICE}), improved sensitivity to DM masses well below the GeV scale is possible by searching for signals 
induced by inelastic processes~\cite{Essig:2011nj}, for which a DM particle is able to deposit much more energy compared to the elastic scattering off of nuclei.  
In particular, one of the most promising avenues is to search for one or a few ionized electrons that are released due to DM particles interacting with \textit{electrons} in the detector~\cite{Essig:2011nj,BRNreport}. 
Indeed, sensitivity to, and resolution of, single or few-electron events has been demonstrated with 
small solid-state targets (silicon and germanium), using data from SENSEI~\cite{Essig:2011nj,Essig:2015cda,skipper2017,sensei2018,sensei2019,SENSEI:2020dpa}, 
SuperCDMS~\cite{Romani:2017iwi,Agnese:2018col,Amaral:2020ryn}, and  EDELWEISS~\cite{Arnaud:2020svb}, as well with noble-liquid (xenon and argon) targets using data from XENON10, XENON100, LUX, XENON1T, PandaX, and DarkSide-50~\cite{Essig:2012yx,Essig:2017kqs,Angle:2011th,Aprile:2016wwo,Agnes:2018oej,LUX:2020vbj,Aprile:2019xxb,PandaX-II:2021nsg,XENON:2021myl}. 

\begin{figure}[b!]
    \centering
      \includegraphics[width=0.48\textwidth]{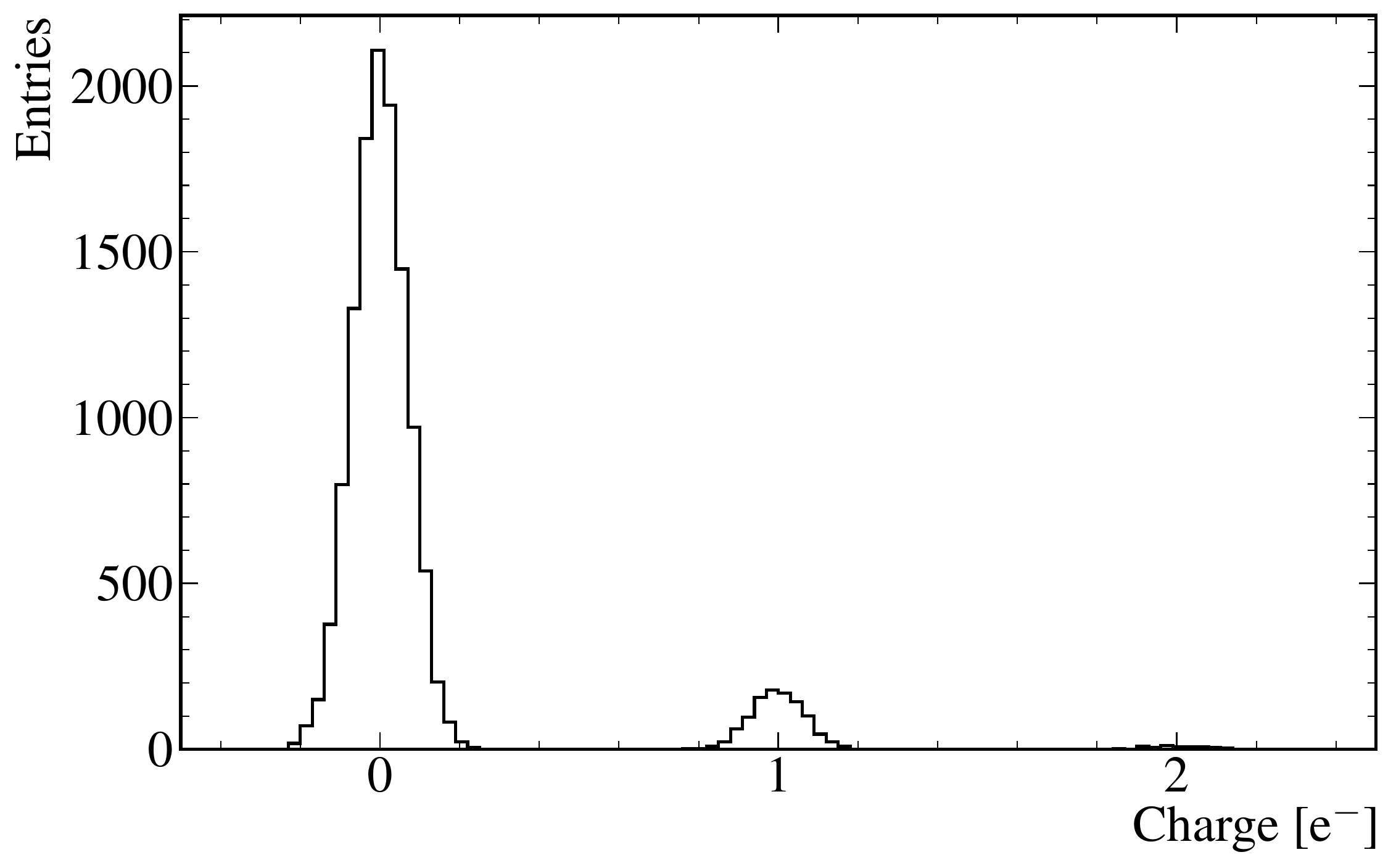}
    \includegraphics[width=0.48\textwidth]{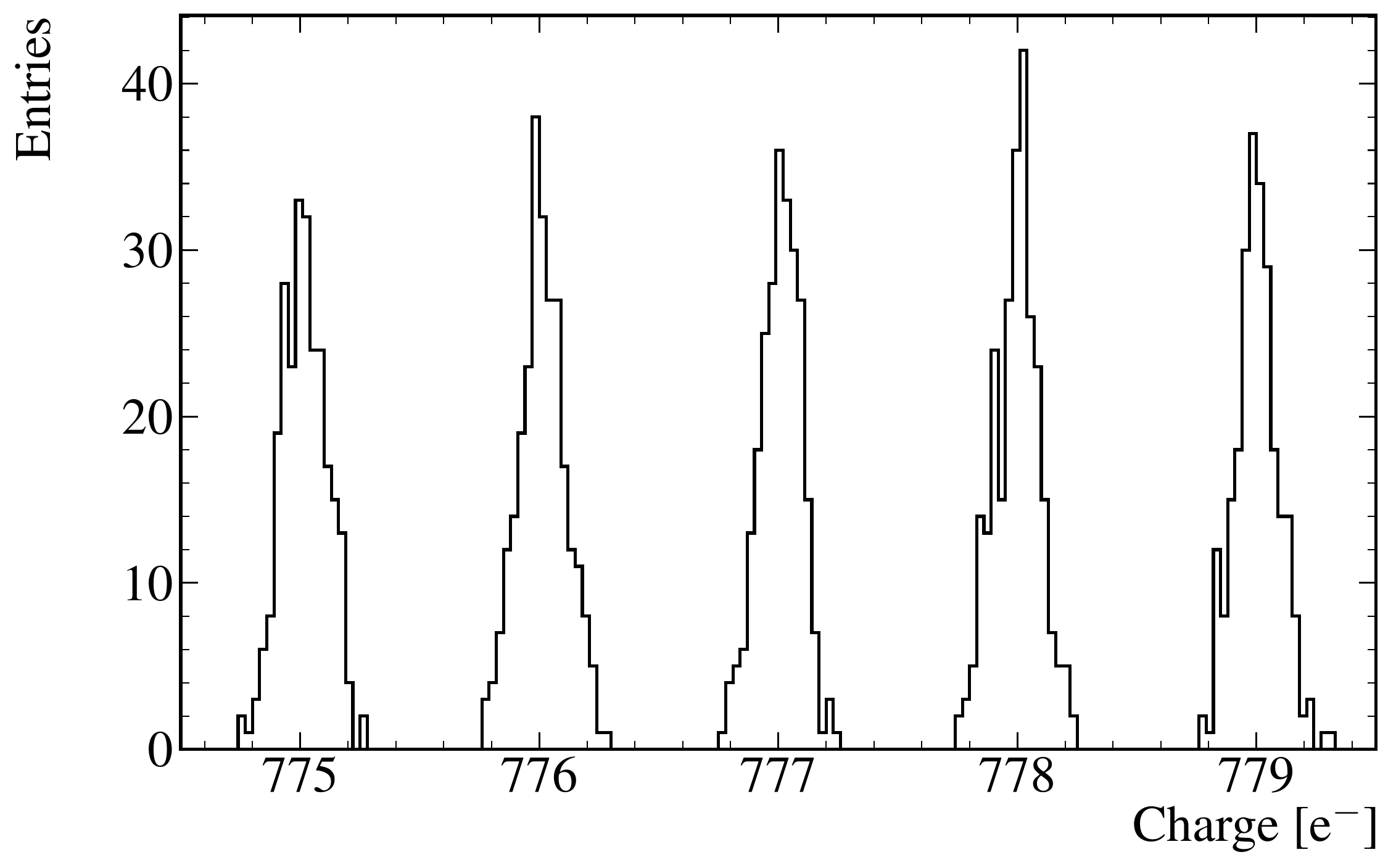}
    \caption{Demonstration of single-electron charge resolution in a skipper-CCD (4000 charge readout samples per pixel; bin width of 0.03\e). The measured pixel charge is shown for pixels with low-light level illumination (left) and stronger illumination (right). Integer electron peaks can be distinctly resolved in both regimes. The peak at 0\e has rms noise of 0.068\e while the peak at 777\e has rms noise of 0.086\e. The two measurements demonstrate the single-electron sensitivity over a large dynamic range. (Figure and caption from~\cite{skipper2017}.) }
    \label{fig:skippernoise}
\end{figure}

\begin{figure}[t!]
    \centering
    \includegraphics[width=0.5\textwidth]{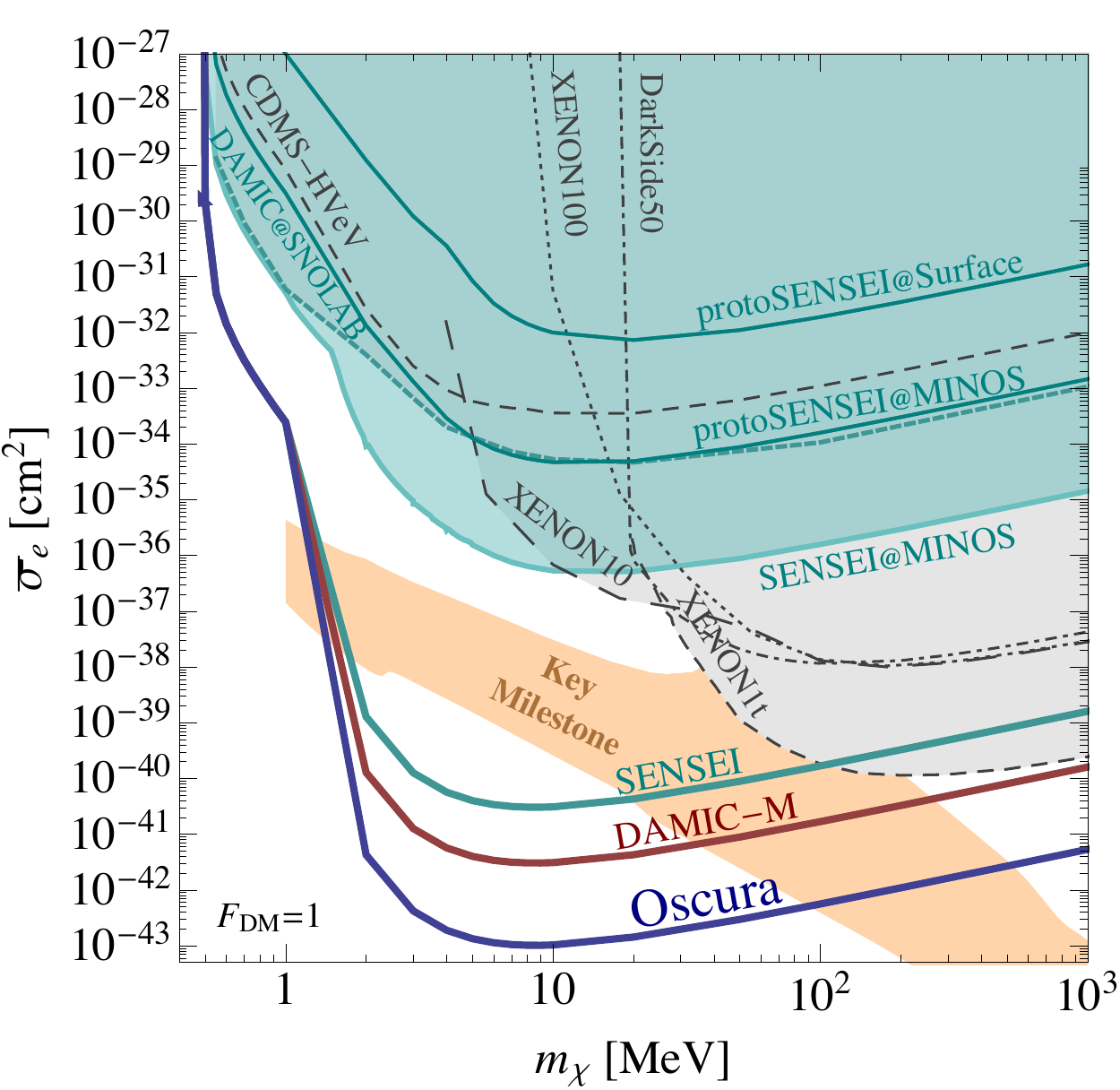}\hfill
    \includegraphics[width=0.5\textwidth]{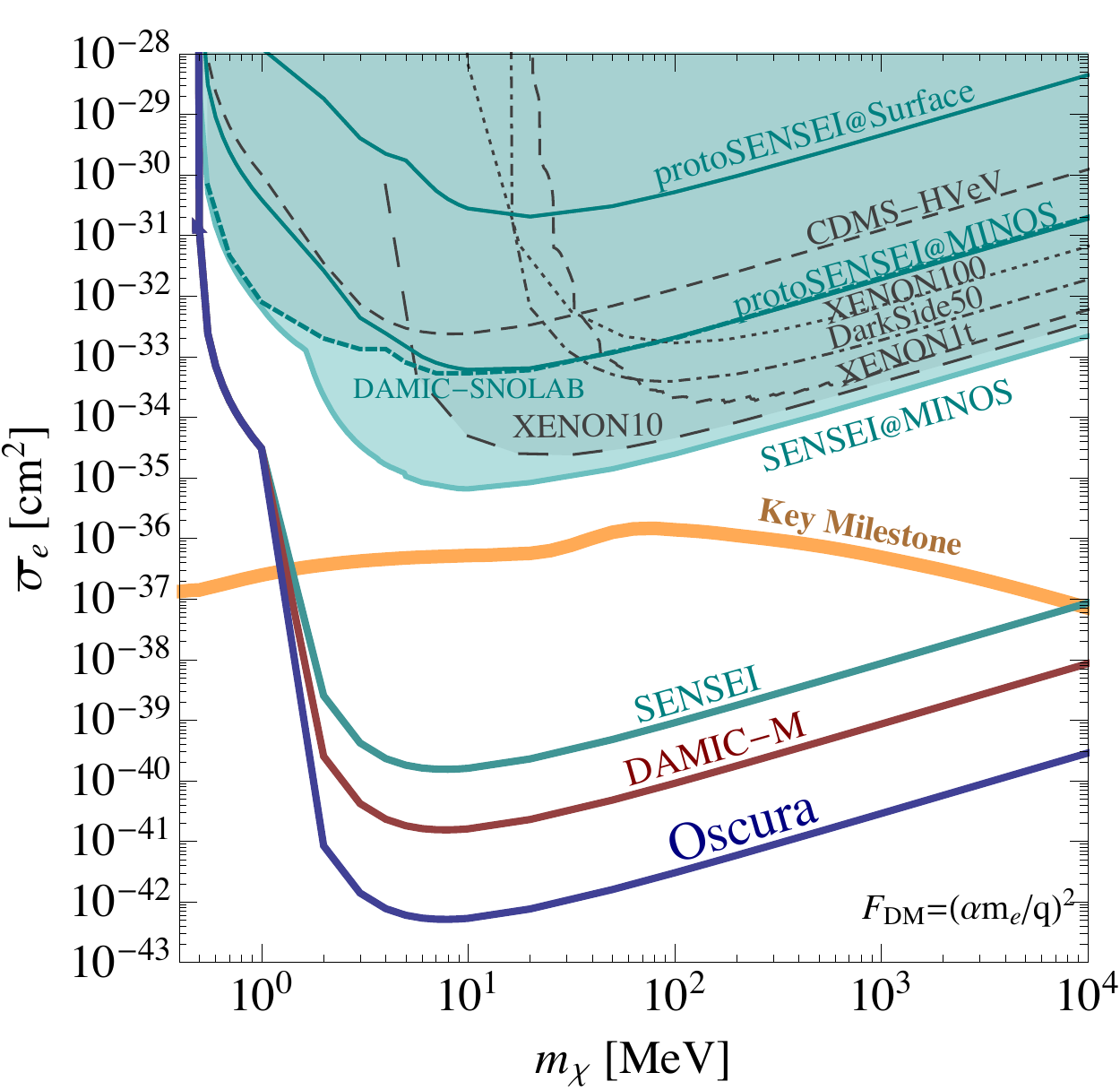}
    \caption{Projected sensitivity to dark-matter-electron scattering of Oscura, assuming a 30 kg-year exposure (blue).  We assume zero background events for events with two or more electrons, and a fixed single-electron ``dark-count'' event rate of $10^{-6} e^-/{\rm pix}/{\rm day}$.  Projected sensitivities for SENSEI (DAMIC-M) are shown in cyan (red)~\cite{Essig:2011nj,Essig:2015cda,skipper2017,Settimo:2018qcm,Castello-Mor:2020jhd,BRNreport}.  Existing constraints from a skipper-CCD from SENSEI are shaded in cyan~\cite{sensei2018,sensei2019,SENSEI:2020dpa}. 
    Shaded gray regions are also constrained by several other experiments (some shown explicitly) including XENON10/100/1t, DarkSide-50, PandaX-II, CDMS-HVeV, and DAMIC at SNOLAB~\cite{Essig:2012yx,Essig:2017kqs,Angle:2011th,Aprile:2016wwo,Aprile:2019xxb,Agnes:2018oej,Agnese:2018col,Aguilar-Arevalo:2019wdi,Essig:2019xkx,Amaral:2020ryn,PandaX-II:2021nsg,XENON:2021myl}.  Orange regions labelled “Key Milestone” are the same as in the BRN report~\cite{BRNreport}.  The left (right) plot assumes the DM-electron interaction is mediated by a heavy (light) mediator.
}
    \label{fig:projection-scattering}
\end{figure}

Among the most promising detector technologies for the construction of a large multi-kg experiment for probing electron recoils from sub-GeV DM are the new generation of silicon Charged Coupled Devices with an ultralow readout noise, so-called ``skipper-CCDs.''  Skipper-CCDs were designed by the Lawrence Berkeley National Laboratory (LBL) Micro Systems Lab.  In 2017, the SENSEI (``Sub-Electron Noise Skipper-CCD Experimental Instrument'') Collaboration demonstrated the ability to measure precisely the number of 
free electrons in each of the million pixels across the CCD~\cite{skipper2017}, see Fig.~\ref{fig:skippernoise}. 
SENSEI packaged a small prototype skipper-CCD sensor and took data at Fermilab on the surface and underground, setting world-leading constraints on DM-electron interactions for DM masses in the range of 500~keV to 5~MeV~\cite{sensei2018,sensei2019,SENSEI:2020dpa}, see Fig.~\ref{fig:projection-scattering}.  
Using new, science-grade skipper-CCDs, pathfinder experiments  
based on this technology are planned for the coming years.  Specifically, SENSEI-100 plans to install a $\sim$100~g detector at SNOLAB during 2022, while DAMIC-M plans to install a $\sim$1~kg detector at Modane during 2023.

\begin{figure}[t!]
\centering
\begin{minipage}[c]{0.55\textwidth}
\includegraphics[width=0.9\textwidth]{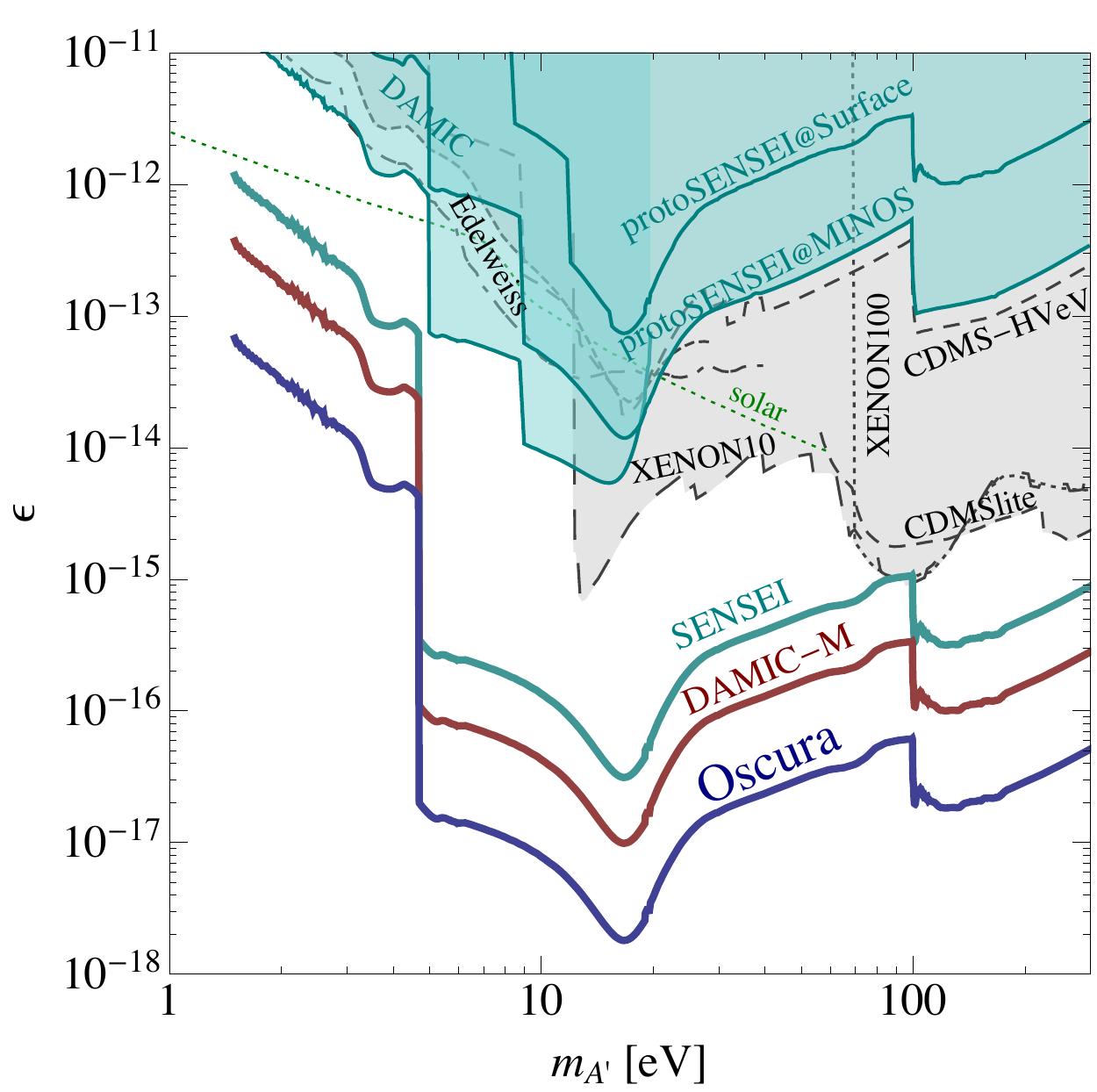}
\end{minipage}\hfill
\begin{minipage}[c]{0.45\textwidth}
\caption{
Projected sensitivity to the absorption of dark-photon DM on electrons for Oscura, assuming a 30 kg-year exposure (blue).  We assume zero background events for events with two or more electrons, and a fixed single-electron ``dark-count'' event rate of $10^{-6} e^-/{\rm pix}/{\rm day}$.  Projected sensitivities for SENSEI (DAMIC-M) are shown in cyan (red). Shaded regions show existing constraints from SENSEI, DAMIC at SNOLAB, XENON10, XENON100, CDMSlite, and CDMS-HVeV ~\cite{sensei2018,sensei2019,SENSEI:2020dpa,DamicHiddenPhoton,An:2014twa,Bloch:2016sjj,Hochberg:2016sqx,Agnese:2015nto,Agnese:2018col,Arnaud:2020svb,An:2013yfc}; the solar constraint is in green~\cite{An:2013yfc}. 
\label{fig:projection-absorption}}
\end{minipage}
\end{figure}

The Oscura experiment will have sensitivity to single electrons, and has the goal of achieving zero background events in the 2--10 electron ionization-signal region (``dark current'', i.e.,  thermal fluctuations of electrons from the valence to the conduction band, presents an irreducible source of single-electron events). Such an experiment will have unprecedented sensitivity to sub-GeV DM that interacts with electrons. In particular, this experiment can probe DM masses in the range of 500~keV to 1~GeV when the DM scatters off electrons through, a ``heavy'' or ``ultralight'' mediator~\cite{Essig:2011nj,Graham:2012su,Essig:2015cda,Essig:2017kqs,Lee:2015qva}, see Fig.~\ref{fig:projection-scattering}.  This would probe many well-motivated sub-GeV DM models that have been highlighted in the recommendations of the Basic Research Needs (BRN) workshop for Dark Matter New Initiatives~\cite{BRNreport}, and which have been combined into the orange region labelled ``Key Milestone'' in Fig.~\ref{fig:projection-scattering} as in the BRN report (see also e.g.~\cite{Essig:2011nj,Essig:2013lka,Alexander:2016aln,Battaglieri:2017aum,BRNreport,Essig:2015cda} and references therein for additional discussions of these benchmark models).  The same detector is also sensitive to bosonic DM that is absorbed by electrons, for DM masses in the range of $\sim$1.1~eV (the silicon band gap) to $\sim$1~keV, see Fig.~\ref{fig:projection-absorption}~\cite{An:2014twa,Bloch:2016sjj,Hochberg:2016sqx}. The projected sensitivities are approximate, and assume 100\% signal efficiency. 

\begin{figure}[htb!]
    \centering
    \includegraphics[width=0.48\textwidth]{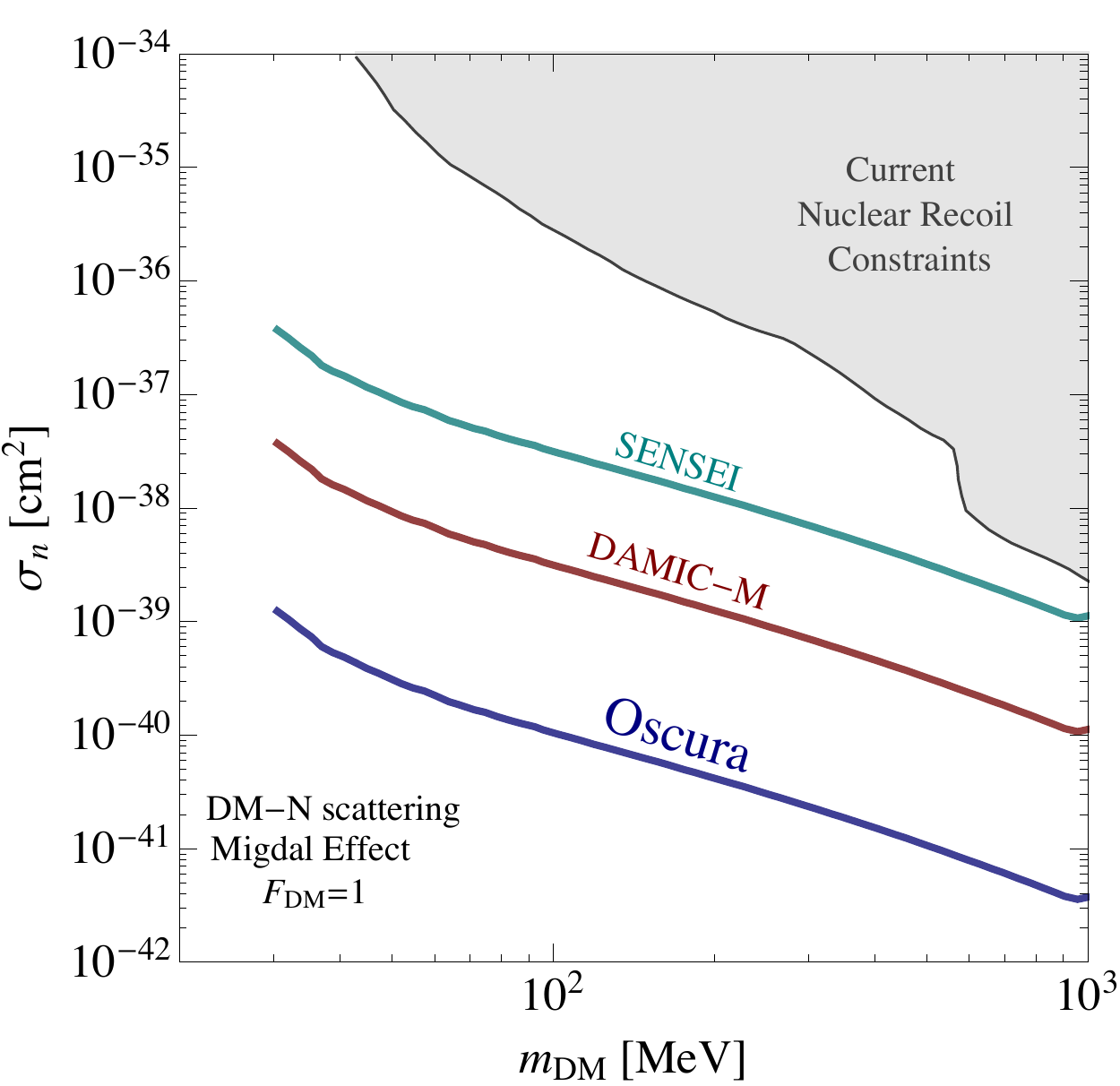}
    \includegraphics[width=0.48\textwidth]{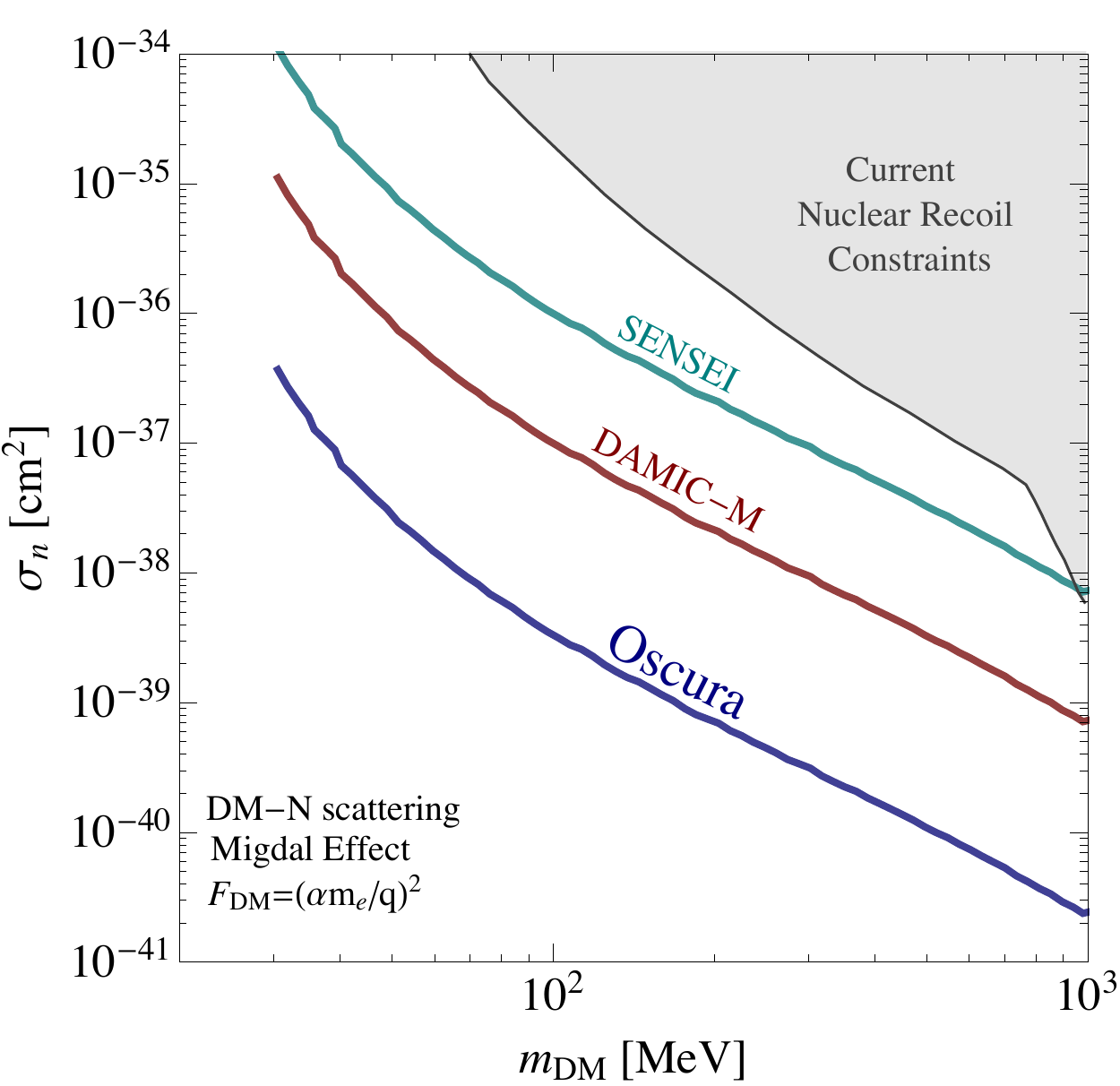}
    \caption{
    Projected sensitivity to dark-matter-nucleus scattering of Oscura, assuming a 30 kg-year exposure (blue).  We assume a two-electron threshold and zero background events. Projected sensitivities for SENSEI (DAMIC-M) are shown in cyan (red).  The projections have been scaled from~\cite{Knapen:2021bwg}. Existing constraints are shaded gray~\cite{Aprile:2019jmx,Akerib:2018hck,Abdelhameed:2019hmk,Liu:2019kzq,Aprile:2019xxb,Essig:2019xkx}. The left (right) plot assumes the DM-nucleus interaction is mediated by a heavy (light) mediator.}
    \label{fig:Migdal}
\end{figure}

The Oscura experiment's ability to measure precisely the number of electrons will allow as well unprecedented sensitivity to DM interacting with \textit{nuclei}, since such interactions can generate one or more ionized electrons in several ways: (i) from the interactions of the recoiling nucleus with the surrounding target material, (ii) from a low-energy photon that is radiated from the nucleus during the DM scattering event, where the photon subsequently converts in the material into electrons~\cite{Kouvaris:2016afs}, and (iii) from the Migdal effect, where the recoiling atom is ionized~\cite{Ibe:2017yqa}.  The latter effect dominates for low-mass DM-nuclear scattering (see e.g.~\cite{Ibe:2017yqa,Bell:2019egg,Dolan:2017xbu,Baxter:2019pnz,Essig:2019xkx,Knapen:2020aky,Knapen:2021bwg}), and would generate multiple electrons, which are easy to measure with the proposed experiment.  We show the expected approximate sensitivity of Oscura to DM-nucleon couplings from the Migdal effect in Fig.~\ref{fig:Migdal}. 

Our choice of silicon skipper-CCDs as the target material, over other possible target materials, is motivated by several important considerations:  
\begin{enumerate}[label=(\roman*),leftmargin=0.2cm,itemindent=.7cm,labelwidth=\itemindent,labelsep=0cm,align=left]\addtolength{\itemsep}{-0.45\baselineskip}
    \item The small silicon band gap allows us to probe DM masses an order of magnitude lower in mass (for both scattering and absorption) than noble-liquid targets, which have an ionization energy of $\mathcal{O}$(10~eV). 
    \item Due to the small silicon band gap, the electron recoil-energy needed to promote additional electrons from the valence to the conduction band is lower in silicon ($\sim$3.8~eV) than in many other target materials,  so that DM-electron scattering events will often contain two or more electrons.
    \item The small mass of the silicon nucleus ensures that solar neutrinos that scatter coherently off nuclei are not a limiting background for the proposed experiment with a 30~kg-year exposure~\cite{Essig:2018tss}. 
    \item The skipper-CCD technology has already been demonstrated and provides an unprecedented charge resolution, extremely low leakage currents, exquisite spatial resolution and three-dimensional reconstruction, and background identification and rejection capabilities.  Indeed, the strongest constraints on low-mass DM scattering off electrons and absorption by electrons (down to $\sim$500~keV and $\sim$1~eV, respectively) are currently obtained with skipper-CCDs~\cite{SENSEI:2020dpa}. 
    \item Rapid progress is being been made in understanding the origin and the mitigation strategies of single- and few-electron backgrounds~\cite{Aguilar-Arevalo:2020oii,Du:2020ldo,DAMIC:2021crr,Proceedings:2022hmu}. 
\end{enumerate}

\section{The Oscura 10 kg DM Detector} \label{sec:techDesc}

The current strategy for an experiment with the exposure and threshold requirements discussed in Table \ref{tab:requirements} is a detector with an active mass of 10 kg using skipper-CCD sensors. Given the standard dimensions of skipper-CCD pixel (15 $\mu$m x 15 $\mu$m) and considering the standard thickness of silicon wafers for processing in commercial foundries (700 $\mu$m), 28 gigapixels are needed to get to an active mass of 10 kg. We have not yet decided on the optimal backside treatment for the  Oscura sensors. For planning purposes we will assume that backside treatment (discussed in Sec.~\ref{sec:backside}) is done at the same foundry where the detectors are fabricated.

The background requirement of less than 1 event between 2e- and 10e- over a 30 kg-year full exposure translates into a background rate requirement of 0.01 dru.\footnote{1~dru (differential rate unit) corresponds to 1~event/kg/day/keV.} This corresponds to a significant improvement over previous CCD experiments, with DAMIC~\cite{DAMIC2016, DAMIC2020} achieving five dru in 2017, SENSEI-100 \cite{sensei2019} designed for five dru, and DAMIC-M~\cite{2020DAMICM} planning for less than one dru. 
This requirement mandates strict control of all materials selected for the experiment and an extensive material assay program, a significant part of the Oscura project plan. 
The background requirement also imposes a cosmogenic activation budget for all parts of the experiment, most significantly for the sensors where we have established a requirement of less than five days of sea level exposure equivalent after tritium removal (see Section~\ref{sec:cosmogenic}).

\begin{table}[H]
	\centering
	\caption{Background requirements.}\label{tab:requirements}
	\small
	\begin{tabular}{cc}
		\hline
		     & requirement \\ \hline
		total exposure   & 30 kg-year \\
		energy threshold & 2e- \\
		background & $<$1 event in full exposure between 2e- and 10e- \\
		\hline
	\end{tabular}
\end{table}

The background requirements in Table~\ref{tab:requirements} also impose an important constraint on the dark count rate for the skipper-CCD sensors. Considering 28 gigapixels and a total exposure time of three years, the dark current rate of $\sim 10^{-6}$ e-/pix/day gives ten events with 2e- considering a Poisson distribution. This number is reduced to a single event with 2e- if the full detector array is readout in less than two hrs, which can be achieved with 24,000 readout channels, each running at 166 pixels/second. 
The operation of skipper-CCDs at this level of dark current requires cooling down the system to somewhere between 120K and 140K (the optimal operating point will be determined from the prototype sensors). The current strategy for the cooling system is to submerge the full detector array in a Liquid Nitrogen (LN2) bath operated with a vapor pressure of 450 psi to reach this temperature. The power for each readout channel is estimated to be 32~mW as measured in the prototype. This corresponds to less than 1~kW power for the full system. The current plan is to provide this cooling capacity with closed-cycle cryocoolers~\cite{AL600}.

The technical requirements are summarized in Table \ref{tab:reqs}.

\begin{table}[H]
	\centering
	\caption{Technical requirements.}\label{tab:reqs}
	\small
	\begin{tabular}{|l|l|l|}
		\hline
		system            & description & goal \\ \hline
		sensor            & readout noise    & 0.15 e- RMS  \\
      	sensor	          & dark current     &   $10^{-6}$ e/pix/day \\
      	readout	          & speed            &   166 pix/sec \\
      	readout           & channel count    &   24,000 \\
		detector array    & total mass & 10 kg \\
		detector array    & number of pixels & 28 Gpix \\
        background        & rate & 0.01 dru \\
        LN2 vessel  	  & operating pressure & 450 psi \\
        cooling   	      & capacity & 1 kW  \\
        DAQ               & data handling & 1 petabyte/year \\
        \hline
		\hline
	\end{tabular}
\end{table}

\subsection{Sensors and Packaging}
\label{sec:packaging}

\begin{figure}[t] 
	\centering 	\includegraphics[width=0.35\linewidth]{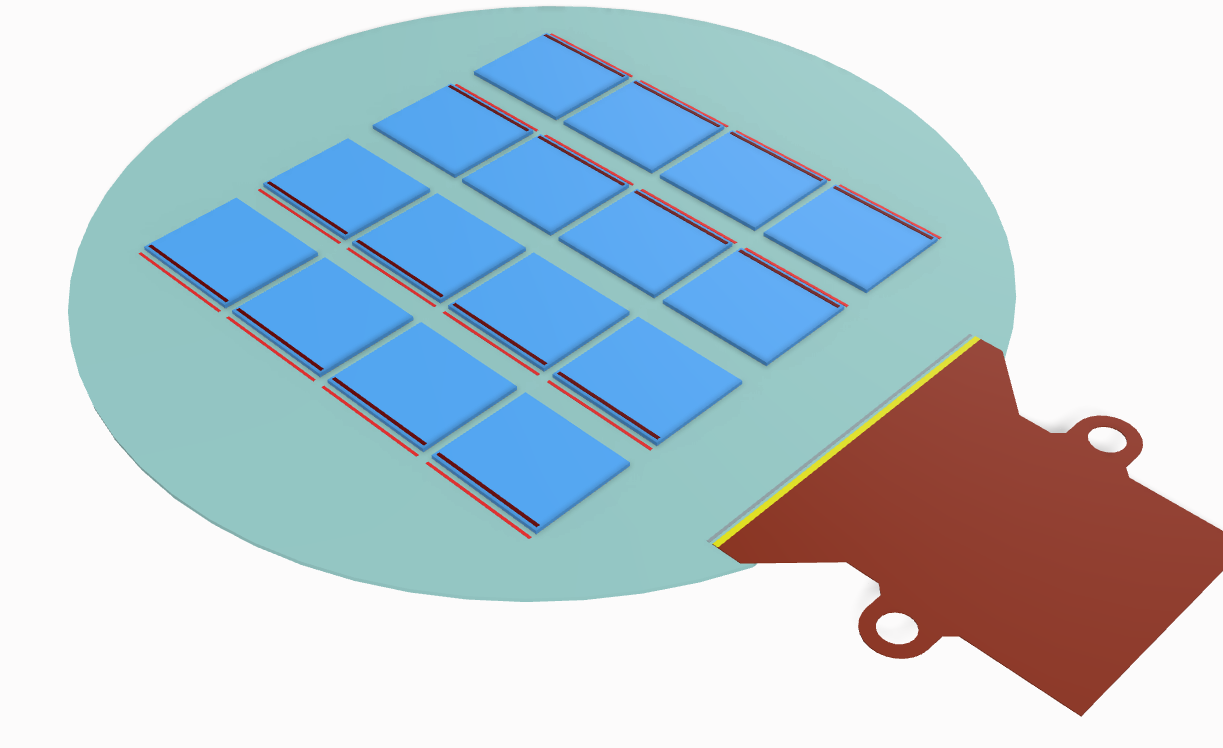} \hfill
	\centering 	\includegraphics[width=0.4\linewidth]{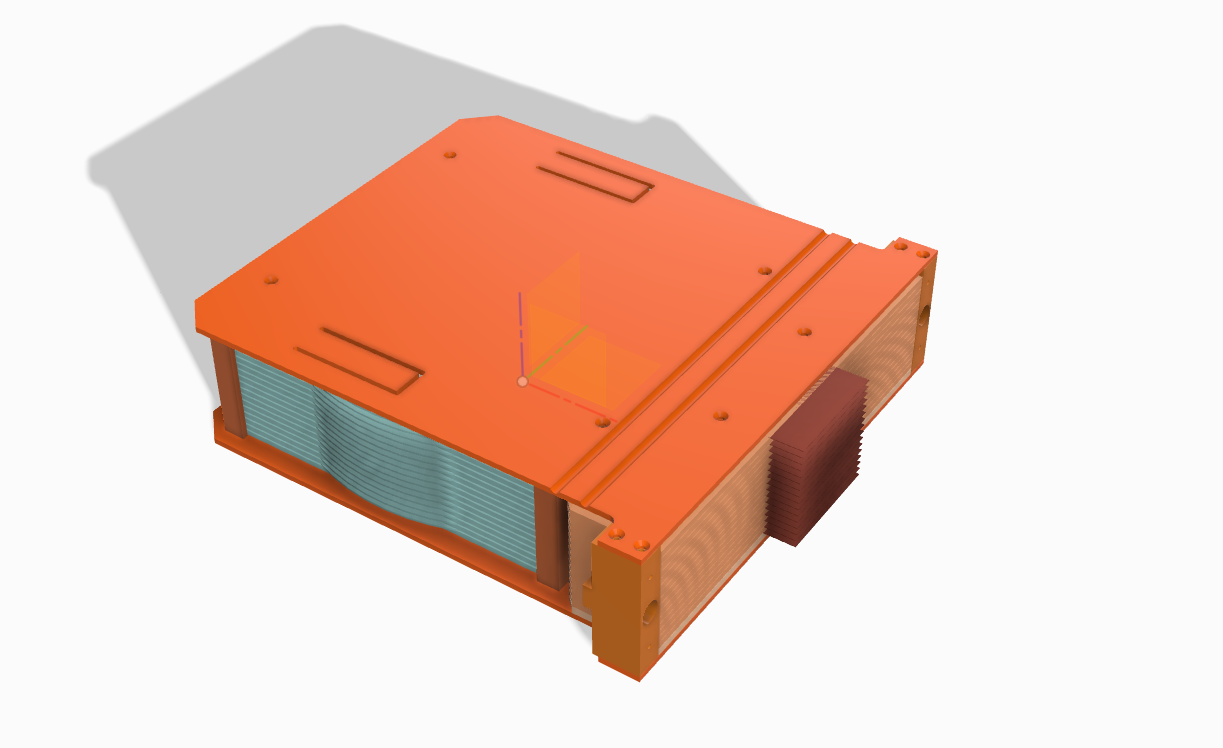}
	\caption{(left) Design of the Oscura Multi Chip Module (MCM) with 16 sensors mounted on a 150~mm silicon wafer. (right) Oscura Super Module (SM) with 16 MCMs supported and shielded with electroformed copper. }
	\label{fig:MCM}
\end{figure}

The packaging of the Oscura sensors will be based on a Multi-Chip-Module (MCM) with 16 sensors epoxied to a 150 mm silicon wafer with traces connecting the CCD to a low radiation background flex as shown in Fig.\ref{fig:MCM}. The package is designed to keep only low-background silicon next to the active volume of the CCD, and a small fraction of the flex consistent with requirements obtained from the simulations (see Section \ref{sec:simulations}). The flex circuit and the sensors are micro-bonded to the carrier silicon wafer. The Oscura experiment needs 1500 MCMs.

As shown in Fig.\ref{fig:MCM}, a Super Module (SM) will be built with 16 MCMs using a support structure of custom ultrapure electro-deposited copper \cite{electroformedcopper}. The SM also includes some copper to shield the radiation of  the first couple of centimeters from the sensors. The Oscura experiment needs 100 SM to reach 10~kg active mass.

\subsection{Liquid Nitrogen Pressure Vessel} \label{sec:vessel}

\begin{figure}[h] 
    \centering
    \begin{minipage}[c]{0.5\textwidth}
    \includegraphics[width=0.97\linewidth]{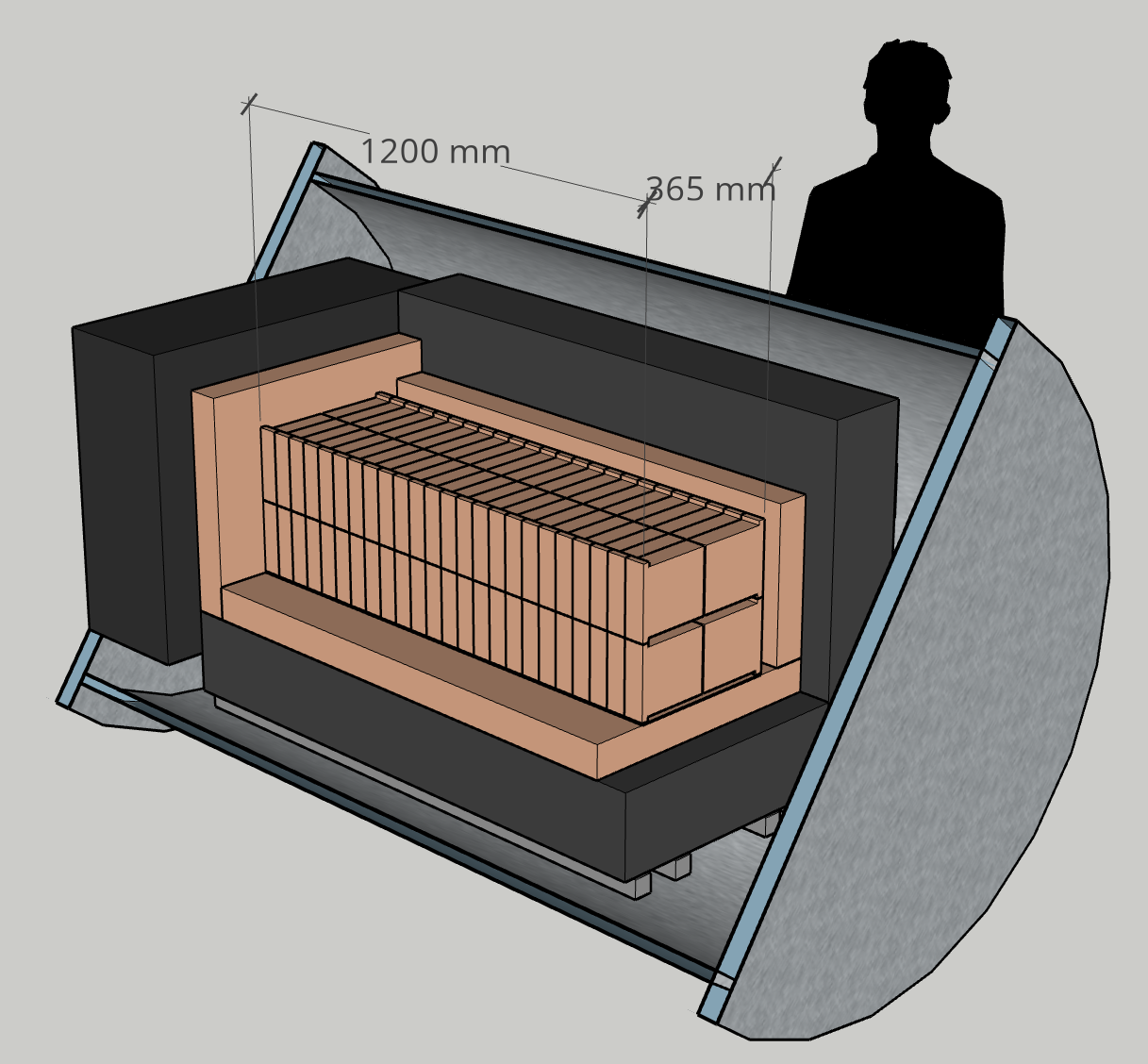}
    \end{minipage}\hfill
    \begin{minipage}[c]{0.5\textwidth}
	\caption{Sketch of a pressure vessel holding the full detector load and an inner lead and copper shields. The dimension lines refer to the ``stacks'' of supermodules, which in this sketch cover 120$\times$36.5~cm. An external shield of $\sim$60~cm of HDPE (or equivalent absorption power of water) and an additional $\sim$10~cm of lead is not shown.}
	\label{fig:10kg2}
	\end{minipage}
\end{figure}

The 10 kg of sensors can be arranged in 100 SMs with dimensions of 120 cm x 50 cm x 50 cm as shown in Fig. \ref{fig:10kg2}. The sensors will operate in an LN2 bath with a vapor pressure of 450 psi. The simulations studies (see Section \ref{sec:simulations}) indicate that a layer of 20 cm of lead is needed between the stainless steel vessel walls and the SMs, and an inner copper shield is necessary to shield the SM from the \isotope{Pb}{210} in the lead. The inner lead and copper shields are also shown in Figure \ref{fig:10kg2}. The total diameter of the pressure vessel is approximately 1~m in this configuration, and some optimization could allow reducing this diameter somewhat. The pressure vessel requires a vacuum jacket to control the thermal load (the vacuum jacket is not shown in the drawings). The cylindrical vacuum jacketed pressure vessel will have access on one of the sides and the detector array with its internal shield assembled outside the vessel and rolled in place. The mechanical design of the pressure vessel is an important part of the engineering effort in the next stage of the R\&D process.

The electronic signals will be carried outside the pressure vessel by using pressure interface boards like those currently under test in the prototype LN2 system \cite{PIBA}. The end cap of the cylindrical pressure vessel has not been designed yet and requires significant engineering effort. This end cap will include the interface board, the cryocooler coldheads to remove the heat out of the LN2 bath, and also ports for filling and venting LN2. The majority of active electronic components will be located between the pressure vessel and lead shield to shield radioactivity in the electronics. 

\subsection{Readout Electronics and DAQ}

The readout electronics for Oscura are divided into Front-End (FE) and Back-End (BE). The FE electronics are formed by the correlated double sampling (CDS) circuit built either with discrete components or with an ASIC as discussed in Section \ref{sec:rdelect}. 

The DAQ will be a cluster of servers on a high-speed network to handle the data and distribute it to facilities at the collaborating institutions for scientific analysis. This will completely be based on off-the-shelf parts and there is no need for development components beyond the commercially available. 

The experiment will have 28 Gpixels running with an exposure time of two hours for a total of three years of science data. This will produce a total of 13,000 science exposures. Assuming 16 bits per pixel, this results in a total of 0.75 petabytes of science data. We will design a DAQ system capable of handling one petabyte of data/year. Data compression schemes are currently under consideration (zero suppression).

\begin{table}[H]
	\centering
	\caption{Readout electronic parts with the counts needed to instrument the full experiment.}\label{tab:eleParts}
	\small
	\begin{tabular}{|l|l|l|l|}
		\hline
		Part  & Description & technology & count needed \\ \hline
		CCDs & sensor & skipper-CCDs & 24,000 \\ \hline
		CDS & analog signal correlated double sampling & ASIC or discrete components& 24,000 \\ \hline
		Mux1 & x16 analog multiplexing for MCM & discrete components on PCB & 1,500 \\ \hline
		Mux2 & x16 analog multiplexing for Supermodule & discrete components on PCB & 100 \\ \hline
        PIBA & carries signal outside the LN2 vessel & high density PCB & 10 \\ \hline
        Osc-LTA & clock drivers and digitization & PCB & 10 \\ \hline
		\hline
	\end{tabular}
\end{table}

\subsection{Radiation Shield}

The external radiation shield for the Oscura experiment has not been designed yet. We expect to use a neutron and gamma shield outside the pressure vessel. The details will be designed in the next two years as part of our development of the Preliminary Design Report for Oscura.

\section{Challenges and Early Technical Studies}
\label{sec:RD}

The technical requirements of the Oscura experiment imposed by the scientific goals discussed above are summarized in Table \ref{tab:reqs}. The Oscura experiment will be composed of an array of skipper-CCD detectors with 28~Gpixels operating at 120K to reduce the dark counts. The Oscura detector concept is described in Section \ref{sec:techDesc}. Three major technical challenges were identified by the project team at the start of the concept design study. The R\&D effort during this stage has been focused on these areas: Sensor development, Front end electronics, and Radiation Background. In this section, we describe the recent process on these high priority R\&D areas.

\subsection{Sensor Development}

\begin{figure}[t] 
    \begin{minipage}[c]{0.4\textwidth}
	\includegraphics[width=0.97\linewidth]{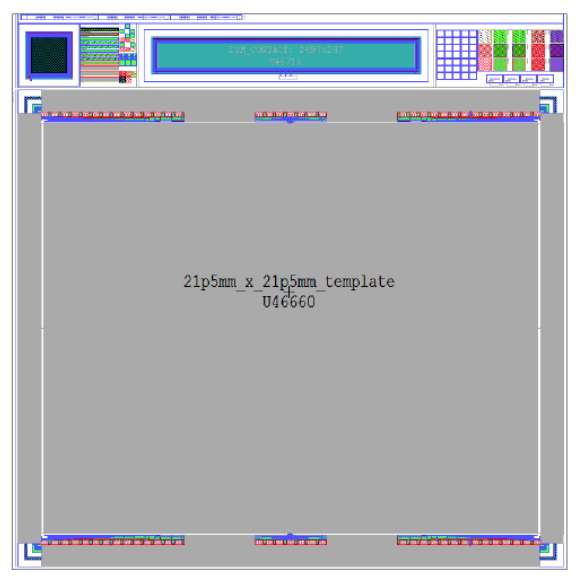}
	\end{minipage}\hfill
	\begin{minipage}[c]{0.6\textwidth}
	\caption{Skipper-CCD sensor design developed for the Oscura experiment to be fabricated at Microchip (design from S.Holland - LBNL). The sensor has four skipper-CCD amplifiers, one on each corner. Because of the step-and-repeat photolithography process used at Microchip, and to avoid the additional challenge of stitching, we are fabricating small format detectors with 1278x1058 pixels and 4 amplifiers. As for previous skipper-CCDs we are using 15 $\mu$m square pixels. The initial run consisted of 24 wafers, and 47 sensors per wafer. The main objective of this fabrication run is to demonstrate the sub-electron noise performance on the new skipper-CCDs from the Microchip foundry. The large quantity of sensors produced during this initial run will be also used to develop prototype packages for Oscura.  }\label{fig:microchip}
	\end{minipage}
\end{figure}

Skipper-CCDs used for existing DM experiments have been fabricated in 150~mm diameter wafers using micro-fabrication tools that are being phased out of modern foundries. A new fabrication process is being developed with Microchip Inc. The new sensors were completed during 2021 and their characterization is ongoing (See Fig.\ref{fig:microchip}).  We have also identified the opportunity to develop more than one fabrication facility (MIT-LL foundry) to improve performance and reduce pressure on the schedule for the experiment construction.

\subsection{Front End Electronics}\label{sec:rdelect}

The 10~kg Oscura experiment will require approximately 24,000 readout channels and a readout time below two hours for the full detector. The requirement for two hour readout is derived from the impact of dark current during the readout time. We are working on the development of a low-cost, scalable readout system for the 10~kg experiment. The main challenge focuses on the front end electronics. In order to avoid having a large number of signal lines going from the sensors (inside the low background cryostat) to the outside world, it is important to develop cold front end electronics and multiplexing. There are two main research thrusts towards developing the readout solution for Oscura. The discrete components solution meets the requirement as discussed in Ref\cite{2021readout}. The ASIC solution could simplify detector assembly (packaging), reduce radioactive background, and reduce cost for the experiment. The current ASIC solutions being considered are the MIDNA\cite{MIDNA} and the repurposing of ASPIC\cite{ASPIC}.

\subsection{Radiation Background}

The Oscura project aims at a radioactive background rate of 10 mdru, i.e. 0.01~counts/kg/keV/day. Achieving this radiation background with an affordable solution constitutes a major risk for the project. We have three R\&D efforts focusing on reducing the risk of the background from components closest to the core of the detector, including the cosmic activation of the silicon detectors. The efforts in understanding the radiation background in Oscura affect directly all other aspects of the experiment design.

\subsubsection{Silicon Activation Studies and Tritium Removal }\label{sec:cosmogenic}

One of the dominant radioactive backgrounds anticipated for Oscura is tritium (\trit). Tritium is a radioactive isotope with a half-life of 12.3 years that undergoes beta decay with a Q-value of 18.6 keV. The continuous spectrum of electrons resulting from \trit~decays leads to a background that spans the energy region of interest for low-mass DM interactions. Tritium and other radioactive isotopes such as \ber~and \sod~are produced in silicon detectors as a result of cosmic-ray exposure, primarily due to interactions of high-energy cosmic-ray neutrons with silicon nuclei in the detector substrate. It is expected that effectively all non-silicon atoms present in the raw silicon material are driven out during single-crystal growth. Consequently, the background level from cosmogenic isotopes in the final detector is effectively determined by the above-ground exposure time during and following detector production, the cosmic-ray flux, and the isotope production cross sections.

The neutron-induced production cross sections for \trit, \ber, and to a lesser extent \sod, were not experimentally known except for a few measurements at specific energies leading to large uncertainties in the expected cosmogenic background. To address this deficiency, several members of the Oscura collaboration performed a dedicated measurement of the integrated isotope-production rates from a high-energy neutron beam. This measurement allowed for a direct extrapolation from the measured beam production rates to the expected sea-level cosmogenic production rates corresponding to ($124 \pm 25$) atoms/(kg day) for $^{3}$H, ($9.4 \pm 2.0$) atoms/(kg day) for $^{7}$Be, and ($49.6 \pm 7.4$) atoms/(kg day) for $^{22}$Na \cite{saldanha2020cosmogenic}. These numbers are in excellent agreement with those indirectly obtained from spectral fits to data in DAMIC at SNOLAB~\cite{DAMIC:2021crr}.

The \trit~production rate corresponds to activity at low energies of roughly 0.002 dru per day of sea-level exposure. This places strong restrictions on the fabrication and transportation of silicon detectors for Oscura. In order to mitigate the tritium background, we are currently exploring the possibility of using elevated temperatures to remove implanted tritium from fabricated silicon devices. It is well-known that at high temperatures hydrogen and tritium are fairly mobile within the silicon lattice and baking can remove them from within silicon substrates \cite{sopori2001silicon}. However, measurements of the tritium diffusion coefficient in silicon vary by several orders of magnitude, thought to be related to variations in the levels of impurities and defects, and the method by which the tritium was implanted (e.g. as a gas or through particle interactions) \cite{sopori2001silicon,ichimiya1968solubility, qaim1978triton, saeki1983origins}.

We have been experimentally measuring the effectiveness of baking to remove tritium produced by high-energy neutron interactions (the dominant cosmogenic production mechanism) from high-purity silicon wafers. Silicon wafers irradiated on a high energy neutron beam at LANSCE \cite{lisowski2006alamos, icehouse} are baked at controlled temperatures to extract the tritium. The tritium is captured, converted to tritiated water, and then measured using liquid scintillation counting \cite{warwick2010effective}. Preliminary measurements have shown that cosmogenically produced tritium can be efficiently removed by baking at $900^{\circ}$C. Detailed results of this study are being prepared for publication.

In addition, we are investigating if tritium can be removed either after fabrication of the CCDs or after packaging. Fabricated CCDs cannot be exposed to very high temperatures and additionally have surface layers (e.g. silicon nitride) that could impede the diffusion of tritium. Similarly, the CCD packaging may not survive very high temperatures. We are currently investigating the maximum temperature that CCDs and CCD packaging can withstand, and the efficiency of tritium removal at lower temperatures (300-500$^{\circ}$C). Early results from baking silicon wafers show promising removal efficiencies at 400$^{\circ}$C and we are putting together a dedicated baking setup for long-duration baking tests at lower temperatures.

\subsubsection{Low Radioactivity Flexible Cable Fabrication}
\label{sec:flexcables}
The baseline design of the Oscura experiment utilizes flat flexible cables, made from copper-polyimide laminates, to readout the CCD sensors. Polyimides are widely used as an insulating substrate in the electronics industry due to their unique properties of high resistivity, high dielectric strength, and flexibility. Polyimides are also stable across a wide range of temperatures, have good thermal conductivity, a thermal expansion coefficient that is close to copper, and a low outgassing rate. 

Commercial laminates of copper and polyimide are typically not very radiopure, with measured contamination levels at roughly 300 parts-per-trillion (ppt) \isotope{U}{238} and 50 ppt \isotope{Th}{232}. However members of the Oscura collaboration have previously worked to identify radiopure laminates with contamination levels at 9 and 20~ppt \isotope{U}{238} and \isotope{Th}{232}, respectively \cite{arnquist2020ultra}.

While radiopure copper-polyimide laminates have been identified, the process of fabricating a cable out of this starting material adds significant contamination. Measurements of completed flexible cables showed some impurities at the level of 2000~ppt \isotope{U}{238} and 800 ppt \isotope{Th}{232}. At these levels, the radioactivity contribution from the flexible cables would dominate the background budget of the entire experiment. We are currently working with a commercial manufacturer of flexible cables to identify and mitigate the sources of contamination during fabrication. Through a careful analysis of each step in the photolithographic production, we have identified specific processes and materials that are the dominant contributors to the contamination and preliminary studies on simplified cables indicate that we can achieve significant reductions in contamination.

\subsubsection{Full Detector Simulations} \label{sec:simulations}

In order to evaluate the background contributions from radioactive contamination in the materials used to construct the experiment, a model is being constructed in GEANT4~\cite{agostinelli2003}.  Although the energy range of interest ($<$1~keV) has not been well-validated in GEANT4, we have adopted the best practices developed by the DAMIC, SENSEI, and SuperCDMS collaborations.  This includes use of the \texttt{EmOption4} high precision low energy electromagnetic physics list, ``Shielding'' physics list for hadronic (neutron) interactions, the \texttt{muelec} low energy silicon list, and tuning low energy atomic deexcitation and secondary production~\cite{valentin2012}.  Continuing to improve and tune the simulation physics list will be a significant ongoing R\&D effort. 
The eventual goal of the simulation effort is to construct a global background budget including all potential radiological background sources. During the initial design development, we have focused on the following sources that we expect to be of greatest impact: 

\paragraph{Cavern background}  We consider specifically the measured backgrounds at SNOLAB: \SI{4000}{neutrons/m^2/day} and \SI{4.25}{gammas/cm^2/s}~\cite{snolabhandbook, picocollaboration2019}. Based on preliminary results, we estimate that a shield of approximately \SI{60}{cm} of high-density polyethylene and \SI{40}{cm} of lead will reduce the flux to acceptable levels. Due to the computationally intense nature of these simulations, this estimate is still in progress. 

\paragraph{Pressure Vessel}    The CCDs will be cooled with liquid nitrogen at $\sim$\SI{425}{psi} in order to raise the boiling point to an acceptable working temperature. This requires a large vacuum-insulated pressure vessel. The baseline design includes a stainless steel pressure vessel and flanges. Properly sourced stainless steel contains radioactive isotopes \isotope{K}{40}, \isotope{U}{238}, \isotope{Th}{232} (KUT) on the order of \SI{1}{mBq/kg} levels with significant variation, and \isotope{Co}{60} typically several times higher than KUT~\cite{akerib2017}. Some fraction of the lead shielding will be placed inside the pressure vessel. Currently, we estimate $\sim 16-20$~cm of lead will be required. Because the cost of the pressure vessel increases rapidly with radius, this value will be carefully optimized.  

We are also considering a titanium vessel in the cost optimization. Titanium has similar levels of KUT contamination per unit mass as steel, but contains almost no \isotope{Co}{60} and has roughly four times the strength-to-weight ratio of steel. So although titanium is much more expensive, it may be possible to use less internal shielding (current estimates indicate 12~cm may be sufficient), thus requiring a smaller vessel.  We have also considered constructing the pressure vessel from oxygen-free high thermal conductivity (OFHC) copper, but at this time it does not appear to be a cost-effective solution. 

\paragraph{Lead shielding}  In addition to KUT, standard lead contains \isotope{Pb}{210} at several hundred \si{Bq/kg}~\cite{Vojtyla:96nimb,alessandrello1991,heusser1991}. The innermost layers of the lead shield will consist of ``ancient'' lead with reduced \isotope{Pb}{210} content.  The exact composition will depend on what grades are available from vendors or for re-use from previous experiments (e.g., DAMIC). As a rough baseline, we consider a 20~cm lead shield with the outer 10~cm standard lead, then 5~cm of lead with 1~Bq/kg \isotope{Pb}{210}, then 5~cm of 0.1~Bq/kg with an 8~cm inner copper layer. Commercial OFHC copper also contains \isotope{Pb}{210} in the bulk at several tens of mBq/kg~\cite{abe2018}, so the innermost centimeters of the copper liner will either have to be very carefully sourced and assayed or be made of electroformed copper. We are in the process of studying what shielding is required for KUT contamination in the lead but expect this to be subdominant. 

\paragraph{CCD modules} In the current design, 16 MCMs (each containing 16 CCDs) are grouped into supermodules, with the full 10~kg experiment containing $\sim$100 supermodules. By placing KUT contaminants at the edges of the MCMs defined in the simulations, we estimate a requirement of $\lesssim$\SI{100}{\nano\becquerel} of \isotope{U}{238} and \isotope{Th}{232} and $\lesssim$\SI{10}{\micro \becquerel} \isotope{K}{40} per supermodule. To meet this requirement, the majority of the mass of the supermodules will be silicon and electroformed copper.  The mass and purity of any epoxies, solder, wirebond material, etc.,  will be carefully considered. The most problematic components will be any non-copper mechanical fasteners (screws, etc) and the readout cables and cold electronics. With proper screening, silicon bronze and Polytetrafluoroethylene (PTFE)~\footnote{PTFE is a synthetic fluoropolymer of tetrafluoroethylene: hydrophobic, non-wetting, high density, and resistant to high temperatures material.}  appear to be suitable mechanical components. Production of the electroformed copper could also benefit from a planned construction of an electroformed copper production facility underground at SNOLAB. 

\paragraph{Cold electronics}  The target radiopurity limits of $\sim$hundreds of nBq per supermodule and the target contamination levels in kapton cables in Section~\ref{sec:flexcables} suggest that only a few centimeters of flex cable can be directly "visible" to the CCDs without some intermediate shielding. The current supermodule design achieves this by clamping the cables between $\sim$\SI{3}{cm} copper shields. Shortly outside the supermodules, the cables will also be routed through the internal lead shield.  The bulk of the cold electronics will be placed outside the lead (next to the pressure vessel).  Further refinement of the simulation will be used to specify the available length of cable, and validate whether any electronic components can be placed inside the lead, e.g., to reduce the un-amplified signal length or provide a connection point to simplify assembly and disassembly. 

\paragraph{Near-infrared (NIR) photon emission, transport, and absorption} High-energy charged particles interacting with the silicon of the CCD and with the surrounding structures can produce NIR photons from Cherenkov radiation. In addition, charged particles produce electron-hole pairs in the highly-doped gettering\footnote{Gettering is the process of reducing impurities by localizing them in predetermined, passive regions of the silicon wafer.} layer on the backside of the CCD that subsequently recombine radiatively and emit near-infrared NIR photons~\cite{Du:2020ldo}.  These NIR photons have a long attenuation length in silicon and, once absorbed, produce a single-electron event. Such events are the likely origin of many of the single-electron events observed in SENSEI data taken at a shallow underground site~\cite{Du:2020ldo,SENSEI:2020dpa}.  Work is ongoing to simulate in detail the production and propagation of these NIR photons.  We expect the single-electron event rate to be reduced drastically in a low-background environment, which we plan to confirm with a detailed simulation. We will also investigate how different thicknesses of the gettering layer on the CCD backside impact the single-electron event rate: while a thick layer enhances the radiative recombination rate, this layer also acts as an efficient absorber of NIR photons generated by Cherenkov. We discuss the CCD backside further in Sec.~\ref{sec:backside}. 

\subsubsection{The CCD Backside}
\label{sec:backside}

The processing of the CCD backside has a non-trivial impact on the observed low-energy background spectrum and rate.  
An important development from the analysis of the latest data from DAMIC at SNOLAB was the discovery of a $\sim$5\,$\mu$m-thick partial charge collection (PCC) region in the backside of the CCDs~\cite{DAMIC2020}.
LBNL CCDs feature a 1\,$\mu$m-thick in-situ doped polysilicon (ISDP) backside gettering layer to remove contaminants from the high-resistivity bulk silicon during fabrication, which also functions as the backside electrical contact of the CCD.
Phosphorous (P) from the highly doped ISDP diffuses with time into the lightly doped high-resistivity silicon.
At intermediate P concentrations, some of the charge generated by ionization events recombines before it diffuses into the fully-depleted region and is drifted across the substrate to the CCD gates.
The loss of charge distorts the spectrum from ionization events on the backside, e.g., surface \pbten\ decays, increasing the amplitude of the background spectrum at low energies, where the DM signal is expected.
Studies of the charge collection efficiency in the backside of LBNL CCDs performed by Oscura collaborators~\cite{PCC2020} confirm the observations at SNOLAB.

As noted previously, the recombination of electron-hole pairs generated in the gettering layer can generate NIR photons which contributes to the single-electron events observed in SENSEI data~\cite{Du:2020ldo,SENSEI:2020dpa}. In addition, a secondary ion mass spectrometry (SIMS) measurement of the DAMIC wafers shows a concentration of hydrogen in the gettering layer of $10^{20}$\,cm$^{-3}$, which implies the presence of radioactive \trit\ on the CCD backside.

LBNL CCDs have been back-thinned in the past to improve quantum efficiency to blue light~\cite{Holland:2003}.
The CCD wafers were sent for backside polishing before the final stages of fabrication to remove all polysilicon (including gettering) and oxide layers.
A backside contact was then formed by depositing a thin ($\sim$20\,nm) ISDP layer.
This process would decrease the thickness of the PCC region by a factor of $10^2$, with a corresponding improvement in the ionization signal lost to recombination by fast electrons (and associated radiative backgrounds) and the \trit\ radioactivity on the backside.  However, this layer also acts as an absorber of some NIR photons from Cherenkov radiation, so its removal may require additional backside treatment to stop external photons.

We are exploring several backside treatment options for the Oscura CCDs and have fabricated and commissioned a calibration system to characterize the response of CCDs to surface backgrounds at the University of Washington.

\section{Oscura Construction }\label{sec:projecplan}

The Oscura construction project has the goal of building the 10 kg detector according to the details in Section \ref{sec:techDesc}. The project is currently in its conceptual design phase which will be completed in 2024. The full 10~kg Oscura experiment construction will be completed in four years after the design phase. Oscura will have a significant detector mass available for scientific studies before the completion of the four years construction project (three kg mass available per year). We are currently exploring the possibility of deploying a fraction of the total mass at an underground facility for early science and background control of the experiment before the construction is completed.

\pagebreak


\pagebreak
\section*{References}
\addcontentsline{toc}{section}{References}
\bibliographystyle{techpubs}
\bibliography{References}

\end{document}